\documentclass[11pt,a4paper]{article}
\pdfoutput=1
\usepackage{jheppub}
\usepackage{graphicx,amssymb,amsmath,amsfonts}
\usepackage[english]{babel}

\makeatother

\def\Duv{{\Delta_{UV}}}
\def\Dir{{\Delta_{IR}}}

\def\lir{{\lambda_{IR}}}
\def\Luv{{L_{UV}}}
\def\Lir{{L_{IR}}}
\def\vev#1{\left\langle #1 \right\rangle}

\title{The holographic dilaton}
\author{Carlos Hoyos, Uri Kol, Jacob Sonnenschein and Shimon Yankielowicz}

\affiliation{School of Physics and Astronomy,\\
The Raymond and Beverly Sackler Faculty of Exact Sciences,\\
Tel Aviv University, Ramat Aviv 69978, Israel}

\emailAdd{choyos, urikol, cobi, shimonya @post.tau.ac.il}

\abstract{
We study a set of examples of holographic duals to theories with spontaneous breaking of conformal invariance in different dimensions. The geometries are domain walls interpolating between two $AdS$ spaces, with a non-trivial background scalar field dual to a relevant operator. We comment on a subtlety in the low momentum expansion pointed out in \cite{Bajc:2013wha} for the case of background gravity and revise the dynamical gravity results of \cite{Hoyos:2012xc}, where the dilaton pole was missing in the scalar-scalar and tensor-tensor two-point functions. We compute the energy-momentum tensor and scalar two-point functions and show that there is indeed a massless dilaton pole.
}


\preprint{TAUP-2969/13}

\begin{document}
\maketitle

\section{Introduction}

Renormalization Group (RG) flows of physical systems between fixed points, which  are very universal phenomena, have been investigated  through the years using various different tools. When the flow is driven by spontaneous breaking of conformal invariance, the conformal anomalies of the UV and IR fixed points should match. As was shown recently by A.~Schwimmer and S.~Theisen \cite{Schwimmer:2010za} this is possible because the dilaton - the Goldstone boson associated to scale transformations - gives a contribution to the anomaly at tree level. This fact has later led to the  proof of the ``$a$ theorem" \cite{Komargodski:2011vj,Komargodski:2011xv}.

In a recent work \cite{Hoyos:2012xc} we studied holographic models with spontaneous breaking of conformal invariance by a relevant scalar operator ${\cal O}$. The breaking produces a flow between two fixed points, that in the holographic description  maps to a domain wall geometry interpolating between two $AdS$ spaces of different radii. A scalar field dual to the relevant operator has a non-trivial background profile in the domain wall geometry.

Using the holographic dictionary we studied correlators of the energy-momentum tensor and the scalar operator at low momentum, expecting to find a dilaton pole. Surprisingly, we found that although the mixed correlator $\vev{T^{\mu\nu}{\cal O}}$ indeed showed the expected behavior,\footnote{The Ward identity of the mixed correlator $\left< T^{\mu}_{\mu}\mathcal{O} \right>$ was also checked in a holographic QCD model with a non-zero gluon condensate \cite{Erdmenger:2011sz}.} a pole was absent in the tensor-tensor and scalar-scalar correlators. In a following paper by Bajc and Lugo \cite{Bajc:2013wha} (see also \cite{Bajc:2012vk} for an earlier work) a similar analysis was made in a simpler setup but with no dynamical gravity. The geometry was fixed to be $AdS$ with a non-trivial background scalar field. They showed that a dilaton pole appears in the scalar-scalar correlator, and pointed out some subtleties of the low momentum expansion. The reason for the appearance of the massless pole is that the normalizable and non-normalizable modes in the bulk mix with each other. As a consequence, each of the modes near the boundary is a superposition of the modes in the near horizon region.

In view of the results of \cite{Bajc:2013wha}, in this note we would like to revisit the low momentum correlators in the domain wall geometries with dynamical gravity for examples where analytic computations can be done explicitly. We will explain the subtlety with the formal low momentum expansion we used in our previous work and will argue using the examples that the appearance of the pole is expected generically, even though we will not give a general proof for arbitrary dimensions.

In two dimensions the analysis is greatly simplified and we will be able to show the appearance of a dilaton pole in general flows. Although a Goldstone mode is in principle not expected on the basis of Coleman's theorem \cite{Coleman:1973ci}, in the large-$N$ limit, implicit in our approach, fluctuations are suppressed and such a mode is possible. We also find that the residue of the dilaton pole is in agreement with anomaly matching arguments. In two spacetime dimensions the kinetic term of the dilaton is proportional to the difference between the central charges of the $UV$ and $IR$ theories $\Delta c$ \cite{Komargodski:2011vj,Komargodski:2011xv}. The two-point function of the dilaton is therefore
\begin{equation}
  \left< \tau\tau \right> \sim \frac{1}{\Delta c}  \frac{1}{q^2}.
\end{equation}
We indeed observe the $1/\Delta c$ factor in the holographic models.

The set of holographic models that we present for dimensions larger than two is a bit special in the sense that these models can be solved analytically, at least in some subsectors. It is not difficult to imagine many more examples of holographic duals to theories with spontaneous breaking of conformal invariance (in fact infinitely many of them). It would be very interesting if some of the models we describe could be embedded in a complete theory, especially because known examples in field theory are relatively scarce. There are two main reasons: (i) Quantum anomalies that break conformal invariance are difficult to avoid and 
(ii) In a conformal field theory, vacua with
nonzero expectation values (VEV) and vanishing energy density are difficult to find.
In supersymmetric  theories the situation is somewhat better, since there are theories where exact flat directions are present also at the quantum moduli space, and in some theories there are    exactly marginal operators.
In non-supersymmetric models such a breaking
could be a non-perturbative effect involving strong interactions. Chiral symmetry breaking in QCD  is  a well known example of dynamical symmetry breaking. However, the analog of the quark anti-quark condensate for the case of conformal invariance is not known.\footnote{ An attempt to construct a holographic model of such a phenomenon was made in \cite{Kuperstein:2008cq} and more examples of this kind are discussed in \cite{BKS}.}

The paper is organized as follows.
In section 2 we review the analysis and results of \cite{Hoyos:2012xc}.
In section 3 we show how the mixing between the normalizable and non-normalizable modes affects the two-point functions of the tensor and scalar modes.
Sections 4-7 are devoted to giving an evidence for the mixing.
In section 4 we study two dimensional theories in the general case.
In section 5 we study an explicit model of a holographic RG flow in arbitrary dimension and solve for the fluctuations.
In section 6 we study a class of examples in various dimensions and explicitly calculate the mixing term.
In sections 7 we study in more details an example in four spacetime dimensions.
We comment on the explicit breaking case in section 8 and conclude in section 9.

\section{Review of previous work}

\subsection{Holographic RG flow}

The dual gravity theory of a $d$-dimensional conformal field theory is an $AdS_{d+1}$ geometry.
An important property of conformal field theories in even dimensions is the presence of conformal anomalies.
The $a$-anomaly coefficient is related to the $AdS$ radius $L$ by \cite{Henningson:1998gx}
\begin{equation}\label{centralCharge}
  a = \frac{ \pi ^{\frac{d}{2}}   }     { \Gamma(\frac{d}{2})       }  \left(  \frac{L}{\ell _p} \right) ^{d-1},
\end{equation}
where $\ell _p$ is the Planck's length.
In two spacetime dimensions the $c$-anomaly coefficient is related to the dual $AdS_3$ radius via the relation \eqref{centralCharge} with the replacement $a\rightarrow \frac{c}{12}$.

The dual gravity theory of a $d$-dimensional RG flow between two fixed points is therefore a geometry that interpolates between two different $AdS$ spaces. The metric that describes this geometry is of the form
\begin{eqnarray}\label{kink}
  ds^2 &=& dr^2+e^{2A(r)}\eta_{\mu\nu}dx^{\mu}dx^{\nu},
\end{eqnarray}
with
\begin{eqnarray}
  \lim_{r\rightarrow\infty}A(r) &=& \frac{r}{\Luv}, \\
  \lim_{r\rightarrow-\infty}A(r) &=& \frac{r}{\Lir}.
\end{eqnarray}
This metric asymptotes to two different $AdS$ spaces with different radii, $\Luv$ and $\Lir$.
The radial coordinate characterizes the RG scale such that the asymptotic $AdS$ space near the boundary is mapped to the $UV$ fixed point and the other $AdS$ space in the interior is mapped to the $IR$ fixed point.
For geometries that satisfy the null energy condition, $A'(r)$ decreases with $r$ and one can construct a monotonically decreasing function along the flow
\begin{equation}
  a(r)=\frac{\pi^{\frac{d}{2}}}{\Gamma\left(\frac{d}{2}\right)\left(\ell_p A'(r)\right)^{d-1}},
\end{equation}
that coincides with the central charges of the theory at the fixed points \eqref{centralCharge}.
This solution is known in the literature as the ``kink geometry'' \cite{Freedman:1999gp} and was studied in the context of the holographic $c$-theorem \cite{Hoyos:2012xc,Freedman:1999gp,Myers:2010tj,Myers:2010xs}.

\begin{figure}
  \center
  \includegraphics[scale=0.85]{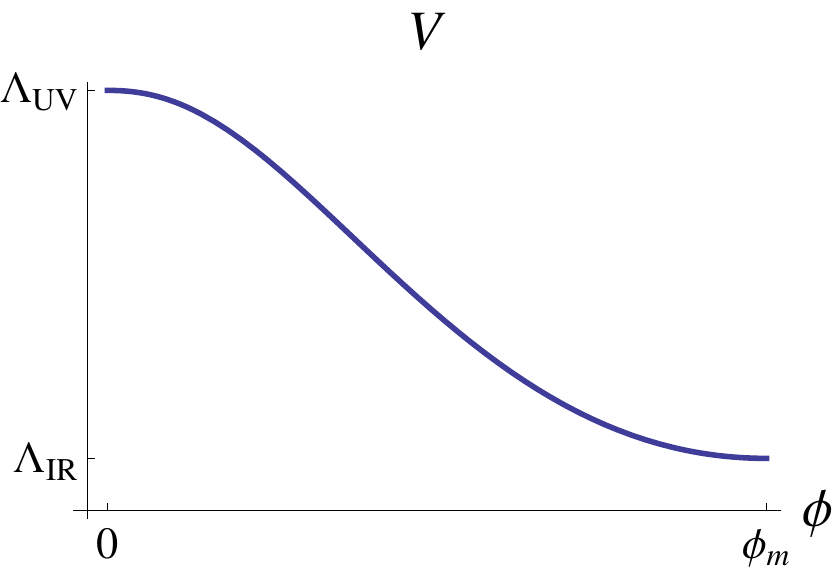}\qquad \includegraphics[scale=0.83]{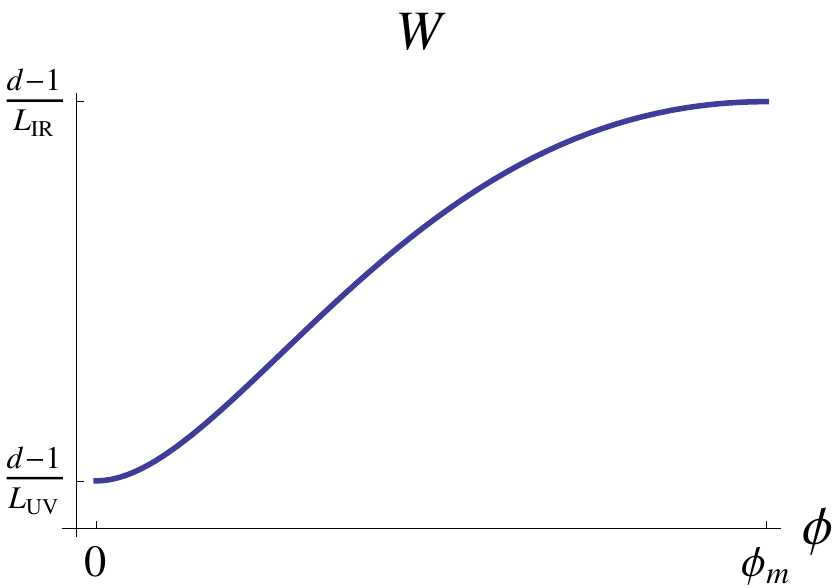}\\
  \caption{The potential and superpotential for the scalar field which corresponds to an RG flow between two fixed points. The critical point at $\phi=0$ corresponds to the boundary, which is dual to the $UV$ fixed point, while the critical point at $\phi=\phi_m$ corresponds to the horizon, which is dual to the $IR$ fixed point. The cosmological constants of the $UV$ and $IR$ $AdS$ spaces are $\Lambda_{UV}=-\frac{d(d-1)}{2\Luv^2}$ and $\Lambda_{IR}=-\frac{d(d-1)}{2\Lir^2}$, respectively.}
	\label{potential}
\end{figure}
The simplest bulk theory for which there is a solution of the form \eqref{kink} is Einstein gravity coupled to a single scalar field with a potential that interpolates between a maximum and a minimum (see figure \ref{potential}). The action is therefore
\begin{equation}
  S = \frac{1}{\kappa^2}\int d^{d+1} x \sqrt{-g} \left( -\frac{1}{2}R-\frac{1}{2}\partial_M \phi \partial^M \phi -V(\phi)\right),
\end{equation}
where $\kappa^2 \equiv (\ell_p)^{d-1}$.
The expansion of the potential around each of the critical points is of the form
\begin{equation}
  V(\phi)=\Lambda+\frac{1}{2}m^2\phi^2 + \dots \quad  ,
\end{equation}
where the cosmological constant is $\Lambda=-\frac{d(d-1)}{2 L^2}$ with $L=\Luv,\Lir$ for the two critical points.
The background solution is homogenous in the field theory coordinates
\begin{equation}
    \phi = \phi_0(r),
\end{equation}
and is monotonic in $r$. In particular, in the vicinity of each of the two $AdS$ spaces its two solutions approach the form
\begin{equation}\label{bckgrdSol}
  \phi_0 \sim e^{-\frac{\lambda}{\Luv} r} , \qquad \lambda= \Delta, d-\Delta.
\end{equation}
$\Delta$ is related to the mass of the scalar
\begin{equation}
  \Delta(\Delta-d)=m^2 L^2,
\end{equation}
and is different at the two critical points. Its value at the maximum of the potential, $\Duv$, is identified as the dimension of the operator in the dual $UV$ theory and its value at the minimum of the potential, $\Dir$, is identified as the dimension of the dual operator in the dual $IR$ theory.
We are interested in flows triggered by relevant deformations ($\Duv<d$). We also impose $\Dir>d$ such that the deformation in the $IR$ is irrelevant and the theory flows to the $IR$ fixed point.
The solution which decays faster near the boundary corresponds to the case where conformal symmetry is broken spontaneously, while the other solution corresponds to explicit breaking.

When the potential can be written in terms of a superpotential \cite{Freedman:2003ax}
\begin{equation}
    V=\frac{1}{2}\left[ (\partial W)^2 - \frac{d}{d-1} W^2 \ \right],
\end{equation}
the equations of motion reduce to a system of first order differential equations
\begin{equation}\label{reducedEOM}
    \phi'=-\partial W  , \qquad A'=\frac{W}{d-1},
\end{equation}
where $'$ denotes a derivative w.r.t. the radial coordinate $r$ and $\partial$ denote a derivative w.r.t. the scalar field $\phi$. Obviously, this system of first order differential equations chooses a particular linear combination of the two solutions \eqref{bckgrdSol}.

It is easy to show that the critical points of the superpotential are also critical points of the potential, so that the superpotential should interpolate two critical points.\footnote{There could be additional critical points of the potential if $\partial^2 W=d/(d-1) W$ for an intermediate value of $\phi$, this does not affect to the analysis presented here.} From its definition, it is clear that $W$ can either be positive or negative. We arbitrarily choose it to be positive, as represented in figure \ref{potential}.
The expansion of the superpotential around each of the critical points is then
\begin{equation}
  W = \frac{d-1}{L} + \frac{\lambda}{2L} \phi ^2 + \dots \quad .
\end{equation}
The $UV$ fixed point is therefore a minimum of the superpotential, since $\lambda_{UV}>0$ for both solutions. The $IR$ fixed point must then be a maximum of the superpotential (otherwise there will be another critical point between the two we discuss) which implies $\lambda_{IR}=d-\Dir$. The choice of $\lambda_{UV}$ selects the asymptotic form of the solution to the first order differential equations. We pick $\lambda_{UV}=\Duv>\frac{d}{2}$, so that the solution asymptotes the subleading mode, which according to the holographic dictionary corresponds to spontaneous rather than explicit breaking of conformal invariance.\footnote{For $\frac{d}{2}-1\leq \Duv\leq \frac{d}{2}$ it is still possible to consider this solution as corresponding to spontaneous breaking, using double-trace deformations \cite{Minces:2001zy}.}

\subsection{Fluctuations}

In \cite{Hoyos:2012xc} we studied low-momentum fluctuations over the classical background solution described above. Let us briefly summarize the analysis and the results.
Fluctuations over the classical background \eqref{kink} can be described in general in the following way \cite{Bianchi:2001kw,Bianchi:2001de}
\begin{eqnarray}
  ds^2 &=& dr^2+e^{2A(r)}(\eta_{\mu\nu}+h_{\mu\nu})dx^{\mu}dx^{\nu}, \\
  \phi &=& \phi_0+\varphi.
\end{eqnarray}
This is called the \emph{radial} or \emph{axial} gauge. The fluctuations $h_{\mu\nu}$ and $\varphi$ are functions of all coordinates.
The metric fluctuation $h_{\mu\nu}$ is decomposed into tensor (transverse and traceless), vector and scalar modes. The vector mode is pure gauge and we will ignore it in the following. The scalar mode is related by the equations of motion to the scalar field fluctuation $\varphi$, so in total there is only one scalar degree of freedom in the bulk. The transverse traceless component of the metric takes the form
\begin{equation}
  h_{\mu\nu}^{TT}=\frac{1}{q^4}\Pi_{\mu\nu}^{\alpha\beta}\epsilon^{\alpha\beta}h^{TT},
\end{equation}
where
\begin{eqnarray}
  \Pi^{\mu\nu,\alpha\beta} &\equiv& P^{\mu\alpha}P^{\nu\beta}-\frac{1}{d-1}P^{\mu\nu}P^{\alpha\beta} ,  \\
  P^{\mu\nu} &\equiv& q^2\eta^{\mu\nu}-q^{\mu}q^{\nu}.
\end{eqnarray}
$P^{\mu\nu}$ is the projection operator onto the transverse component, $q^{\mu}$ is the $d$-momentum and $\epsilon^{\alpha\beta}$ is an arbitrary symmetric tensor. In the rest of this paper we will omit the $^{TT}$ superscript which stands for transverse traceless. We will refer to the tensor fluctuation simply as $h$. We will also define $Q\equiv\sqrt{-q^2}$.

It will be convenient to use the background solution as the radial coordinate. That can be done since the background solution is monotonic in $r$.
The scale factor, for instance, can be written as a function of $\phi_0$
\begin{eqnarray}
  A &=& -\frac{1}{d-1} \int d\phi_0 \frac{W}{\partial W}.
\end{eqnarray}

The equation of motion for the scalar fluctuation $\varphi$ is a third order ordinary differential equation. However, we can bring it to a simpler form using the following change of variables
\begin{equation}
  \left( \partial - \frac{\partial^2 W}{\partial W} \right) \varphi = \frac{W}{(\partial W)^2}e^{-dA}G .
\end{equation}
$G$ is a gauge invariant variable and its equation of motion is a second order ordinary differential equation. It is possible to write the action for the fluctuation directly in terms of a gauge invariant variable, and derive its equation of motion \cite{Kofman:2004tk,Berg:2005pd,Elander:2010wd}, but we keep the notations of our previous paper \cite{Hoyos:2012xc} in order to make the comparison easier.
The equation of motion for the scalar, as well as for the tensor, mode is then of the form
\begin{eqnarray}\label{EOM}
    \partial^2 f + \partial B \partial f +\frac{Q^2e^{-2A}}{(\partial W)^2} f &=& 0 ,
\end{eqnarray}
where $f$ is either the scalar variable $G$ or the tensor fluctuation $h$. $B$ is a function of the background fields
\begin{eqnarray}
  \label{tensorB} B^{(h)} &=& \ln \partial W +d A , \\
  \label{scalarB} B^{(G)} &=& 2\ln W- \ln \partial W  -(d-2)A .
\end{eqnarray}

Near the asymptotic boundary of the space we can solve for the fluctuations. There are two solutions
\begin{eqnarray}
  \label{BoundarySeries1}
  \varphi &=& \varphi_b \phi_0 ^{\frac{d}{\lambda_{UV}}-1}     + \varphi_{n}\phi_0  +\dots ,
   \\
   \label{BoundarySeries2}
  h &=& h_{b}+h_{n} \phi_0^{\frac{d}{\lambda_{UV}}}+\dots .
\end{eqnarray}
`$\dots$' refers to subleading corrections to each of the solutions. The coefficient of the leading solution near the boundary is identified with the source for the dual operator while the coefficient of the subleading solution is identified with its VEV. $h_{b}$ is therefore identified as the source for the dual energy-momentum tensor. For the scalar the identification depends on the value of $\lambda_{UV}$. For $\lambda_{UV}=\Duv>\frac{d}{2}$, corresponding to the spontaneous breaking case, $\varphi_{b}$ is identified as the source for the dual scalar operator, while for $\lambda_{UV}=d-\Duv<\frac{d}{2}$, corresponding to the explicit breaking case, $\varphi_{n}$ is identified with the source.

\subsection{Boundary conditions}

In order to determine the solution uniquely the ratio $\frac{\varphi_n}{\varphi_b}$ should be fixed.\footnote{More precisely, we should determine the functional dependence of $\varphi_n$ on $\varphi_b$, but for the purpose of computing two-point correlation functions the linear dependence is enough.} This is done using the boundary conditions on the horizon. Equation \eqref{EOM}, for both the tensor and scalar sectors, can be solved order by order in momentum. The solution is composed of two independent series
\begin{equation}\label{pertSol2}
    f_{pert}= \sum_{n=0}^{\infty} \left[ D_0 U_{n}(\phi_0) +D_1 V_{n}(\phi_0) \right] Q^{2n},
\end{equation}
where
\begin{eqnarray}
  \label{u0}
  U_0 &=& 1 ,\\
  \label{v0}
  V_0 &=& \int d\phi_0 e^{-B},
\end{eqnarray}
and
\begin{equation}
  F_{n+1} = -  \int  d\phi_0 e^{-B}   \int  d\phi_0 \frac{  e^{B-2A}     }  { (\partial W)^2  } F_n .
\end{equation}
$F_n$ stands for either $U_n$ and $V_n$.
The coefficients $D_0$ and $D_1$ are related to the coefficients in the expansions \eqref{BoundarySeries1}-\eqref{BoundarySeries2} as follows
\begin{eqnarray}
  \label{transCoeff1} Tensor:& \qquad D_0 =h_b  ,& \qquad  D_1=h_n ,\\
  \label{transCoeff2} Scalar:& \qquad D_0 =\frac{\left(d-2 \Duv\right) \Duv }{(d-1) \Luv}\varphi _b  ,& \qquad  D_1=  -\frac{2(d-1)}{\Duv}Q^2  \varphi_n.
\end{eqnarray}
The small expansion parameter of this solution is really $Q^2 e^{-2A}$. For arbitrarily small momentum this perturbative solution extends deep into the bulk. However, for a given momentum the expansion will always break down close to the horizon at $r\rightarrow-\infty$ where the factor $e^{-2A}$ diverges. For this reason, one cannot impose boundary conditions directly on the perturbative solution, since it is not valid on the horizon.

The procedure for imposing boundary conditions on the perturbative solution is called the \emph{matching procedure}. The equation of motion for the fluctuation \eqref{EOM} can be solved exactly in the near horizon region. For the second order differential equation there are in general two possible solutions
\begin{equation}\label{nearHSol}
  f_{horizon} = C_0 H^{(1)}\left(\Lir Q e^{-2A}\right) + C_1 H^{(2)} \left(\Lir Q e^{-2A}\right),
\end{equation}
which depend on the dimensionless parameter $\Lir Q e^{-2A}$. Imposing boundary conditions amounts to fixing the ratio $\frac{C_0}{C_1}$.
For sufficiently low momentum the perturbative solution \eqref{pertSol2} extends all the way to the near horizon region (but never reaches the horizon itself) such that there is a region of space where both solutions are valid. This overlapping region is defined to be where $\Lir Q e^{-2A} \rightarrow 0 $ and the geometry is well approximated by the near horizon geometry, simultaneously. Evaluating the perturbative solution \eqref{pertSol2} in the near horizon region should therefore be equal to taking the limit $\Lir Q e^{-2A} \rightarrow 0$ of the near horizon solution \eqref{nearHSol}. Matching the two solutions in the overlapping region, such that this equality holds and the solution is consistent, determines the perturbative solution uniquely.
The matching procedure hence provides a tool to transfer the information about the boundary conditions in the $IR$ to the asymptotic boundary where the dual field theory lives.

In order to match the perturbative solution \eqref{pertSol2} with the near horizon solution \eqref{nearHSol} one needs to evaluate the functions $U_n,V_n$ in the near horizon region. However, these functions are expressed as integrals over functions of the background fields and, in general, without specifying the exact form of the background, one cannot solve them. Therefore, in \cite{Hoyos:2012xc} we first expanded the integrands around the horizon and then performed the integrals.
The result of this procedure (for both $UV$ and $IR$ fixed points) is
\begin{equation}\label{nhv}
    V_0 \sim \phi_0 ^{\alpha},
\end{equation}
and
\begin{equation}
    \frac{U_{n+1}}{U_n} \sim \frac{V_{n+1}}{V_n} \sim \phi_0^{\frac{2}{\lambda}},
\end{equation}
where
\begin{equation}
  \alpha=\left(   \frac{d}{\lambda},\frac{2\lambda -d+2}{\lambda}   \right),
\end{equation}
for the tensor and scalar cases, respectively.
In the near horizon region, on the other hand, the equation of motion \eqref{EOM} takes the form
\begin{eqnarray}
    \partial^2 f + \frac{1-\alpha_{IR}}{\phi_0-\phi_m} \partial f +  \left( \frac{\Lir Q}{\lir} \right)^2 (\phi_0-\phi_m)^{\frac{2}{\lir  }-2} f  &=& 0.
\end{eqnarray}
The solutions for this equation are the two Hankel functions
\begin{eqnarray}
  f_{horizon} &=&  y ^{ \nu }
  \left[
  C_0 H_{\nu}^{(1)}  \left( y \right)
  +
  C_1 H_{\nu}^{(2)}   \left( y \right)
  \right]
   ,\\
   y&\equiv& \Lir Q (\phi_o-\phi_m)^{\frac{1}{\lir}}
    ,\\
  \nu &=& \frac{\alpha_{IR} \lir}{2}.
\end{eqnarray}
Imposing the ingoing boundary conditions for this solution \cite{Son:2002sd} chooses one of the Hankel functions $C_0=1,C_1=0$. Expanding the near horizon solution at low momentum now gives
\begin{equation}
  f_{horizon} \sim  1+ P(Q) \phi_0 ^{\alpha_{IR}},
\end{equation}
with
\begin{equation}
    P(Q) \sim \frac{1}{\Lir}(\Lir Q)^{d}\ln (\Lir Q)^2 , \frac{1}{\Lir}(\Lir Q)^{2\lir-d+2},
\end{equation}
for the tensor and scalar cases, respectively. The $\ln (\Lir Q)^2$ term in the tensor case is present only in even dimensions $d$.
Matching with the perturbative solution then fixes the ratio between the two solutions
\begin{equation}\label{naiveBC}
    \frac{D_1}{D_0} = P(Q).
\end{equation}

\subsection{Correlators}

To calculate two-point functions in the dual field theory one has to expand the on-shell bulk action to second order in the fluctuations. The result contains divergent terms that have to be cancelled using appropriate local counterterms. The finite part of the action is then
\begin{eqnarray}\label{action2nd}
  \nonumber S_{finite} &=&     \frac{1}{8 \kappa^2 \Luv}   \int
  \frac{d^d q}{(2\pi)^d}
  \left[
    \left (  \frac{1}{q^4}\Pi^{\mu\nu,\alpha\beta}  \frac{d \: h_n}{h_b}    \right)         h_{b,\mu\nu} h_{b,\alpha\beta}
       \right.
   \\
  && \left.
   \qquad\qquad \qquad\qquad\qquad+
   M^{\mu\nu}(q)
   \varphi_b h_{b,\mu\nu}
   + 4(2\Duv-d) \varphi_b \varphi_n
   \right],
\end{eqnarray}
with
\begin{equation}
  M^{\mu\nu}(q)=-\frac{2\Duv(2\Duv-d)}{d-1}\frac{1}{q^2}P^{\mu\nu}+4(2\Duv-d)\eta^{\mu\nu}.
\end{equation}
The function $M^{\mu\nu}(q)$ is purely kinematic and completely independent of the boundary conditions.
$h_{b,\mu\nu}$ is the source for the energy-momentum tensor and $\varphi_b$ is related to the source of the scalar field $J_b$ as follows \cite{Hoyos:2012xc}
\begin{equation}
  \varphi_b = \frac{\kappa^2 \Luv}{2\Duv-d} \left< \mathcal{O} \right> J_b.
\end{equation}
Differentiating the action \eqref{action2nd} twice with respect to the different sources, $J_b$ and $h_{b,\mu\nu}$, will now result in different two-point functions. The two-point function of the mixed correlator is
\begin{eqnarray}\label{mixedCorrelator}
  \left< T^{\mu\nu}\mathcal{O} \right> &=& \frac{1}{8 } \frac{\left< \mathcal{O} \right>}{2\Duv-d}    M^{\mu\nu}(q)
  =      -\frac{\Duv\left< \mathcal{O} \right>}{4(d-1)}\frac{1}{q^2}P^{\mu\nu}+  \frac{1}{2}\left< \mathcal{O} \right>\eta^{\mu\nu}.
\end{eqnarray}
There is a massless dilaton pole, in agreement with Goldstone's theorem.
The tensor-tensor and scalar-scalar correlators are
\begin{eqnarray}
  \label{correlators1} \left< T^{\mu\nu}T^{\alpha\beta}   \right>  &=&      \frac{d}{8 \kappa^2 \Luv} \frac{1}{q^4}\Pi^{\mu\nu,\alpha\beta}  \frac{ h_n}{h_b} ,\\
  \label{correlators2} \left< \mathcal{O}\mathcal{O} \right> &=&  \frac{ \kappa^2 \Luv }{2\Duv-d}   \left< \mathcal{O} \right>^2  \frac{\varphi_n}{\varphi_b}.
\end{eqnarray}

Using \eqref{transCoeff1}-\eqref{transCoeff2} and the boundary conditions \eqref{naiveBC} we arrive at the final result
\begin{equation}\label{tt}
  \left< T^{\mu\nu}T^{\alpha\beta}   \right>  \sim    \frac{1}{Q^4}\Pi^{\mu\nu,\alpha\beta}  (\Lir Q)^{d}\ln (\Lir Q)^2,
\end{equation}
for the tensor-tensor correlator and
\begin{eqnarray}\label{oo}
  \left< \mathcal{O}\mathcal{O} \right> &\sim&   (\Lir Q)^{2\lir-d} =  (\Lir Q)^{d-2\Dir},
\end{eqnarray}
for the scalar-scalar correlator. The scalar-scalar correlator is divergent at zero momentum as expected from Goldstone's theorem, but a massless dilaton pole is missing in both tensor-tensor and scalar-scalar correlators, which is quite surprising. As we will explain in the next section there is actually a subtlety in the formal expansions we have used, the expressions above will be affected. As a result, the dilaton pole appears as suggested by \cite{Bajc:2013wha}.


\section{The mixing}\label{mixing}

In \cite{Bajc:2013wha}, a simplified version of the system described in the previous section was studied - a scalar field with a potential interpolating between two critical points on a fixed $AdS$ background. The simplified example revealed a possible subtle mistake in our analysis which can be seen when a concrete example is studied. The analysis of \cite{Bajc:2013wha} showed that the near horizon limit of the function $V_0$ \eqref{v0}, when calculated explicitly, may be different than the result when first taking the limit of the integrand and then performing the integral. More precisely, the authors of \cite{Bajc:2013wha} found that in the near horizon region the function $V_0$ qualitatively behaves as
\begin{equation}
    V_0 \sim 1+ \phi_0^{\alpha},
\end{equation}
(instead of \eqref{nhv}) and therefore there is a mixing between the two series. More generally, different terms in the expansion \eqref{pertSol2} may mix with each other when evaluated at the near horizon region.

Our main purpose in this paper is to study explicit examples for the theories studied in \cite{Hoyos:2012xc} and check if there is a mixing between the normalizable and non-normalizable terms. Note that a mixing is expected to exist on general grounds. In the low momentum expansion we select a basis of linearly independent solutions based on the asymptotic behaviour close to the boundary. The solutions computed in the near horizon region form a basis as well, but in general a different one. The mixing between solutions simply reflects this fact. We cannot completely discard that for some suitably chosen superpotentials the two basis are aligned, since that will require finding exact solutions in arbitrary dimensions.

In this section we show how the mixing affects the boundary conditions and rederive the results for the different correlators. If different terms in the expansion \eqref{pertSol2} mix when expanding around the horizon, the boundary conditions \eqref{naiveBC} will change.
For instance, if there is a constant term in $V_0$ when expanding around the $IR$ fixed point, it will mix with the constant solution of the first series. Such mixings will modify the boundary conditions \eqref{naiveBC} to
\begin{equation}\label{newBC}
    \frac{D_1 + c_1 (Q^2+\dots) D_0}{D_0 + c_2 (1+\dots)D_1} = P(Q),
\end{equation}
where $(\dots)$ refers to higher order corrections in momentum. Solving \eqref{newBC} for the ratio between the two coefficients we get
\begin{equation}\label{newRatio}
    \frac{D_1}{D_0} = \frac{P(Q) -c_1 (Q^2 +\dots)}{1-c_2P(Q)(1+\dots)}.
\end{equation}
For $P(Q) \sim Q^n$ we can than expand \eqref{newRatio} for three different cases:
\begin{enumerate}
  \item $n<0$: \quad\qquad$\frac{D_1}{D_0} \sim -\frac{1}{c_2} + Q^2+\dots+\frac{1}{P(Q)}+\dots$,
  \item $0<n<2$:\qquad $\frac{D_1}{D_0} \sim P(Q)(1+Q^2+\dots)$,
  \item $n>2$: \quad\qquad$\frac{D_1}{D_0} \sim c_1 Q^2 (1+Q^2+\dots+\frac{1}{Q^2} P(Q)+\dots)$.
\end{enumerate}
The tensor mode corresponds to the last line and the scalar to the first one.

Using the results for the tensor and scalar two-point functions \eqref{correlators1}-\eqref{correlators2}, and \eqref{transCoeff1}-\eqref{transCoeff2}, we conclude that
\begin{eqnarray}
  \label{correlatorsMixing1} \left< T^{\mu\nu} T^{\alpha\beta} \right> &\sim& \Pi^{\mu\nu,\alpha\beta} \frac{1}{Q^2}(c_1+Q^2+\dots+Q^{d-2}\ln Q^2+\dots) ,\\
  \label{correlatorsMixing2}  \left< \mathcal{O}\mathcal{O} \right> &\sim&  \frac{1}{Q^2} (-\frac{1}{c_2} + Q^2+\dots+Q^{d-2-2\lir}+\dots).
\end{eqnarray}
The result for the mixed correlator $\left< T^{\mu\nu}\mathcal{O} \right>$ \eqref{mixedCorrelator} does not change since it is purely kinematic and is independent of the boundary conditions.
At low momentum equations \eqref{correlatorsMixing1}-\eqref{correlatorsMixing2} are equivalent to
\begin{eqnarray}
\label{CorrSponT}
  \left< T^{\mu\nu} T^{\alpha\beta} \right> &\sim& \Pi^{\mu\nu,\alpha\beta} \frac{1}{Q^2+\dots+Q^{d}\ln Q^2+\dots} ,\\
\label{CorrSponS}
    \left< \mathcal{O}\mathcal{O} \right> &\sim&  \frac{1}{Q^2+\dots+Q^{2\Delta_{IR}-d}+\dots}.
\end{eqnarray}
These are the right correlators for a dilaton interacting with a CFT. In particular, the term $\sim Q^{2\Delta_{IR}-d}$ is proportional to the coupling of the dilaton to the leading irrelevant operator of the $IR$ CFT. Note that in the limit of very low momentum we are left with just the dilaton pole, which reflects the decoupling of the dilaton from the CFT.
We discuss and interpret these results in more details in the conclusions section.

\section{Exact solution in 2D}\label{exact2D}

In two spacetime dimensions ($d=2$) it is possible to solve the scalar equation \eqref{EOM} at zero momentum exactly. The reason is that when $d=2$ the function $B^{(G)}$ \eqref{scalarB} gets extremely simplified since the coefficient in front of the scale factor $A$ vanishes. $B^{(G)}$ then reads
\begin{equation}
  B^{(G)} = -\ln\left[\frac{\partial W}{4W^2}\right]   =  -\ln \left[-\partial\left(\frac{1}{4W} \right)\right].
\end{equation}
We have included a normalization factor inside the $\log$ to match with the normalization of the dilaton later on.
The non trivial solution to the equation of motion \eqref{v0} can then be computed exactly
\begin{equation}\label{2dSol}
  V_0 = \int d\phi_0 e^{-B} =  -\int d\phi_0 \partial\left(\frac{1}{4W} \right) = -\frac{1}{4W}.
\end{equation}
The expansion of \eqref{2dSol} around each of the fixed points is then
\begin{equation}
   V_0 =  - \frac{L}{4}  \left( 1-\frac{\lambda}{2} \phi_0^2 + \mathcal{O}(\phi_0^3)  \right).
\end{equation}
We see that both expansions around the $UV$ and $IR$ fixed points contain a constant piece. The constant piece in the expansion around the $UV$ fixed point will mix with $U_0$ and will be at odds with the expansion of the solution for $\varphi$ around the boundary \eqref{pertSol2}. We therefore need to subtract it. After subtracting this constant, the expansion around the $IR$ fixed point contains a non trivial constant piece which will mix with the other solution. The mixing term is therefore
\begin{equation}
  c_2 = \frac{\Luv -\Lir}{4}.
\end{equation}
Using \eqref{transCoeff2}, \eqref{correlators2} and the fact that to leading order $\frac{D_1}{D_0}=-\frac{1}{c_2}$,
the leading behavior of the scalar-scalar two-point function at small momentum is hence
\begin{equation}\label{2dScalarCorr}
  \left< \mathcal{O}\mathcal{O} \right> =  - 2 \left( \Duv\left< \mathcal{O} \right> \right) ^2  \frac{\kappa^2 }{\Luv-\Lir} \frac{1}{Q^2}.
\end{equation}

This result agrees with anomaly matching arguments in field theory. When conformal symmetry is spontaneously broken, the total central charge $c_{UV}$ of the theory remains constant along the RG flow \cite{Schwimmer:2010za}. However, in the $IR$ the degrees of freedom split between a (possibly strongly coupled) $IR$ CFT of central charge $c_{IR}<c_{UV}$ and a dilaton. The dilaton contributes to the central charge at tree level, compensating for the difference. This can be seen directly from the Wess-Zumino action \cite{Komargodski:2011vj,Komargodski:2011xv}
\begin{equation}
  S_{dilaton}  =  \frac{\Delta c}{24\pi} \int d^2x \sqrt{g} \left(  \tau R + (\partial \tau)^2 \right),
\end{equation}
where $\Delta c = c_{UV}-c_{IR}$ is the difference between the $UV$ and the $IR$ central charges. The dilaton's two-point function in two flat spacetime dimensions is therefore
\begin{equation}
  \left< \tau \tau \right>  =  - \frac{24\pi}{\Delta c}\frac{1}{ Q^2},
\end{equation}
(recall that $Q^2\equiv -q^2$).
We will now compare to our result using the holographic dictionary \eqref{centralCharge}
\begin{equation}
  c = \frac{12 \pi L}{\kappa^2}.
\end{equation}
We then expect that the two point function computed in holography takes the form
\begin{equation}\label{dilaton}
  \left< \tau \tau \right>  =  -2 \frac{\kappa^2}{\Luv-\Lir}\frac{1}{ Q^2}.
\end{equation}
In order to compare \eqref{dilaton} to \eqref{2dScalarCorr} we need to take into account the overlap between the operator $\mathcal{O}$ and the dilaton $\tau$ \cite{Schwimmer:2010za}
\begin{equation}
  \mathcal{O}  \simeq  e^{\Duv\tau} \left<  \mathcal{O}  \right>  =  \left<  \mathcal{O}  \right> + \Duv  \left< \mathcal{O} \right>  \tau  +  \dots \quad.
\end{equation}
The relation between the $\mathcal{O}$-correlator and the $\tau$-correlator is therefore
\begin{equation}\label{relation}
  \left< \mathcal{O}\mathcal{O} \right> = \left( \Duv  \left< \mathcal{O} \right>  \right)^2 \left< \tau\tau \right> + \dots \quad .
\end{equation}
The solution we found \eqref{2dScalarCorr} and the expectation from field theory \eqref{dilaton} indeed obey this relation.

Note that even though there is still a pole in the scalar-scalar correlator, in two dimensions the leading singularity of the tensor-tensor two-point function \eqref{correlatorsMixing1} is
\begin{equation}
  \left< T^{\mu\nu} T^{\alpha\beta} \right> \sim \Pi^{\mu\nu,\alpha\beta} \frac{\ln (\Lir Q)^2}{Q^2}.
\end{equation}
The reason is that the contributions form the $IR$ CFT are dominant at low momentum over the contributions from the dilaton. This does not happen for the scalar-scalar correlator because the contributions from the $IR$ CFT come from an irrelevant operator.

\section{Holographic RG flow model}\label{model}

In the previous section we solved the zero momentum scalar equation in two dimensions exactly. More generally, in higher dimensions and for the tensor equation, it is not possible to do so without specifying the precise form of the superpotential.
In order to solve the zero momentum equation in the more general case we will study in this section a particular model of an holographic RG flow.

It would be convenient to change variables to
\begin{eqnarray}
  v &\equiv& e^{\beta \phi_0}.
\end{eqnarray}
The potential is then
\begin{eqnarray}
  V &=& \frac{1}{2} \left[ \beta ^2 v^2 (\partial_v W) ^2 -\frac{d}{d-1} W^2  \right].
\end{eqnarray}
The equations of motion for the tensor and scalar modes are
\begin{eqnarray}\label{eqV}
   \partial_v ^2 f + \partial_v \tilde{B}  \partial_v  f  + \frac{Q^2e^{-2A}}{\left[(\beta v)^2\partial_v W\right]^2} f &=& 0,
\end{eqnarray}
where
\begin{eqnarray}
    \tilde{B}^{(h)} &=& 2\ln v +\ln\partial_v W +d A,\\
\label{BGfun} \tilde{B}^{(G)} &=&  2\ln W -\ln \partial_v W -(d-2)A,
\end{eqnarray}
and
\begin{equation}
      A = -\frac{1}{(d-1)\beta^2} \int dv \frac{W}{v^2\partial_v W}.
\end{equation}

We consider a superpotential interpolating between two critical points, as depicted in figure \ref{potential}. The derivative of the superpotential is therefore of the form
\begin{eqnarray}
  \partial_v W &=& -\frac{1}{L_0}\frac{(v-1)(v-b)}{v^c},
\end{eqnarray}
where $1<v<b$. $v=1$ is a minimum of the superpotential and corresponds to the $UV$ fixed point while $v=b$ is a maximum of the superpotential and corresponds to the $IR$ fixed point.
$L_0$ has dimensions of length and was introduced in order to account for the dimensions of the superpotential.
The superpotential is therefore given by
\begin{eqnarray}\label{superpotential}
  W &=& -\frac{1}{L_0} v^{1-c} \left( -  \frac{b}{c-1}  +\frac{b+1}{c-2} v  - \frac{1}{c-3} v^2  \right).
\end{eqnarray}
With this choice of superpotential the dimensions of the dual operators in the $UV$ and $IR$ are
\begin{eqnarray}
  \label{Duv} \Duv &=& -(d-1)\beta ^2\frac{(b-1) (c-3) (c-2) (c-1)}{b (c-3)-(c-1)} ,\\
  \label{Dir} \Dir &=& d+  (d-1) \beta ^2\frac{(b-1) (c-3) (c-2) (c-1) }{b (c-1)-(c-3)} ,
\end{eqnarray}
and the $AdS$ radii are
\begin{eqnarray}
  \frac{\Luv}{L_0} &=& -(d-1)\frac{(c-3) (c-2) (c-1) }{b (c-3)-(c-1)} ,\\
  \frac{\Lir}{L_0} &=& (d-1)b^{c-2}  \frac{ (c-3) (c-2) (c-1) }{b (c-1)-(c-3)}.
\end{eqnarray}

This model is parameterized by three variables $(\beta,b,c)$. A more physical parametrization would be to use $\Duv,\Dir$ instead of two of them. We find it convenient to use the following set of variables to describe the system
\begin{equation}
    (\beta,b,c)\rightarrow(\Duv,\Dir,c).
\end{equation}
One can invert \eqref{Duv}-\eqref{Dir} to express $\beta,b$ in terms of the new set of variables

\begin{eqnarray}
\label{btrans}
  b &=& \frac{    (c-3) ( \Dir-d) +(c-1) \Duv        }{       (c-1) ( \Dir-d)+(c-3) \Duv      },\\
\label{betatrans}
    \beta^2 &=&    \frac{   2\Duv(\Dir-d) } {  (c-3)(c-1) (d-1) (d + \Duv- \Dir )   }.
\end{eqnarray}

In the rest of this paper we will set the dimensionful parameter of the model to one $L_0=1$. All the dimensionful quantities are measured in units of $L_0$.

\subsection{The tensor mode}

With the superpotential \eqref{superpotential} the equation of motion of the tensor mode at zero momentum is
\begin{eqnarray}\label{eqtensor}
  \partial_v ^2 h +  \omega \frac{(v-\gamma )(v-\tau )}{v(v-1)(v-b)}   \partial_v  h &=& 0 .
\end{eqnarray}
$\omega,\gamma$ and $\tau$ depend on the three parameters of the model and are listed in appendix \ref{ExactSol}.
Equation \eqref{eqtensor} has an exact solution
\begin{equation}\label{tensorSol}
h=v^{a_1} F_1\left[a_1,a_2,a_3,a_1+1,\frac{v}{b},v\right],
\end{equation}
where $F_1$ is the Appell hypergeometric function of the first kind and
\begin{eqnarray}
  a_1 &\equiv& 1-\frac{\gamma  \tau  \omega }{b} = (c-1)-(c-3) \frac{ d (d-\Dir+\Duv)}{2 (\Dir-d) \Duv} ,\\
  a_2 &\equiv& \frac{(b-\gamma ) (b-\tau ) \omega }{(-1+b) b} = \frac{\Dir}{\Dir-d}  ,\\
  a_3 &\equiv& \frac{(-1+\gamma +\tau -\gamma  \tau ) \omega }{-1+b} = 1-\frac{d}{\Duv}.
\end{eqnarray}
This solution corresponds to $V_0$ in the expansion \eqref{pertSol2}. Our aim is to check whether this solution mix with the constant solution.

\subsection{The scalar mode}
The scalar equation is more complicated due to the $\ln W$ term in the function $\tilde{B}^{(G)}$ \eqref{BGfun}.
We can bring it to a form similar to the tensor equation using the following change of variables
\begin{equation}
  G(v) = X(v) Y(v).
\end{equation}
The equation of motion for $Y(v)$ at zero momentum is then
\begin{equation}\label{equationY}
  \partial_v^2 Y(v) + \left[\tilde{B}^{(G)}+2\ln X(v) \right] \partial_v Y(v) + \left[ \frac{\partial_v\tilde{B}^{(G)}\partial_v X(b)+\partial_v^2X(v)}{X(v)} \right]  Y(v) = 0.
\end{equation}
Choosing
\begin{equation}
  X(v)\sim \frac{1}{W(v)},
\end{equation}
then cancels the $\ln W$ term.

In the present case we choose
\begin{equation}
    X(v)=\frac{v^{1-c}}{W(v)}.
\end{equation}
With the superpotential \eqref{superpotential} the equation for $Y(v)$ then takes the form
\begin{equation}\label{scalarEq}
    \partial_v ^2 Y + \omega \frac{(v-\gamma )(v-\tau )}{v(v-1)(v-b)}  \partial_v  Y    -  \frac{2(\omega+1)(v-\alpha)}{v(v-1)(v-b)}  Y = 0,
\end{equation}
where the parameters $\omega,\gamma$ and $\tau$ take different forms than the tensor case (see appendix \ref{ExactSol}).
Equation \eqref{scalarEq} has an exact solution\footnote{We used Maple to solve equation \eqref{scalarEq}.}
\begin{equation}\label{scalarSol}
    Y=H\left[ b, -2(\omega+1)\alpha,-2,\omega+1,j_1,j_2,v  \right],
\end{equation}
where H is the Heun-G function and
\begin{equation}
  \omega =  (d-2) \frac{(c-1) (d-\Dir+\Duv)}{2 (d-\Dir) \Duv}  -c,
\end{equation}

\begin{equation}
\alpha = \frac{
(\Dir-d)^2\left[(d-2)(c-3)-2 \Duv (c-2) \right]
+\left[  d(c-1)  -2(c-2)\Dir   +2(c-3)  \right] \Duv^2
}
{\left[(c-1)(d - \Dir) -(c-3)\Duv\right]
\left[
(d- 2)(d-\Dir)         -(d+2-2\Dir)\Duv
\right]},
\end{equation}

\begin{eqnarray}
  \nonumber j_1 &\equiv& \frac{\gamma \omega \tau}{b} ,\\
  &=&   \frac{(c-3) (d-2) (d-\Dir+\Duv)}{2 (d-\Dir) \Duv} -(c-2)  ,\\
 \nonumber j_2 &\equiv& -\frac{\omega (\gamma  \tau -\gamma -\tau +1)}{b-1}, \\
   &=& \frac{d-2-\Duv}{\Duv}.
\end{eqnarray}
For brevity, we omit the values of $\gamma$ and $\tau$, which are listed in appendix \ref{ExactSol}.

\section{Solutions in various dimensions}
The Appell hypergeometric and the Heun functions are expressed as a series expansions around the singular points of the equation ($v=0,1,b,\infty$). However, their global properties are not known and therefore we cannot use them to check whether there is a mixing between the normalizable and non-normalizable modes. In order to do that we need to reduce the Appel and Heun functions to simpler functions (like the hypergeometric function) for which the global properties are known and the mixing can be checked. Our strategy will be to tune the c-parameter in such a way that the complicated functions will reduce to simpler functions. The operators' dimensions $\Duv,\Dir$ are in principle not fixed, but for some range of parameters they may not necessarily be compatible with the tuning of c. We will show that there is a range of parameters for $\Duv,\Dir$ which is both compatible with the tuning of c and the requirements on the potential.

\subsection{The tensor mode}\label{TensorSolVarD}
The Appell hypergeometric function \eqref{tensorSol} can be reduced the the incomplete Euler beta function when
\begin{equation}\label{condition1}
    a_1+1=a_2+a_3.
\end{equation}
That happens when c is tuned to
\begin{equation}\label{cTunedT}
    c=\frac{d^2-d \Dir+5 d \Duv-4 \Dir \Duv}{d^2-d \Dir+3 d \Duv-2 \Dir \Duv}.
\end{equation}
Then the solution for the tensor fluctuation is reduced to
\begin{equation}\label{reducedTensorSol}
    h=B\left(x;a_1,1-a_2\right),
\end{equation}
where
\begin{equation}
x\equiv \frac{b-1}{b}\frac{v}{v-1}.
\end{equation}

The incomplete Euler beta function can be expanded around each of the fixed points. Its expansion around $v=1$ ($x=\infty$) is
\begin{equation}\label{tbe}
  h = \frac{(-1)^{-a_1} \Gamma(1+a_1) \Gamma(a_2-a_1)}{a_1 \Gamma(a_2)} + x^{a_1-a_2} \left[    \frac{(-1)^{-a_2}}{a_1-a_2}+\frac{(-1)^{-a_2} a_2}{(-1+a_1-a_2) x}  +\mathcal{O}\left(x^{-2}\right)    \right].
\end{equation}
Its expansion around $v=b$ ($x=1$) is
\begin{equation}\label{the}
    h=\frac{\pi  \text{Csc}\left(\pi  a_2\right) \Gamma \left(a_1\right)}{\Gamma \left(1+a_1-a_2\right) \Gamma \left(a_2\right)}+
     (x-1)^{1-a_2}  \left[      \frac{1 }{a_2-1}+ \mathcal{O}(x-1)       \right].
\end{equation}

The expansion of the solution around the boundary \eqref{tbe} includes a constant piece which is not part of the usual expansion of the solution in $AdS$. Therefore we need to subtract it from the solution. In the expansion around the $IR$ fixed point we then get a constant piece which mix with the constant solution
\begin{eqnarray}\label{mixCoeffTensor}
     c_2
    &=&  \frac{(-1)^{-a_1} \pi  \Gamma(a_1) \left[\text{Csc} (\pi  \left(a_1-a_2\right))+(-1)^{a_1} \text{Csc} (\pi  a_2)\right]}{   \Gamma(a_2)  \Gamma  \left(1+a_1-a_2\right)}.
\end{eqnarray}
This expression will vanish only if both $a_1$ and $a_2$ are integers and $a_1>0$. To check this possibility we express $\Duv$ and $\Dir$ in terms of $a_1$ and $a_2$
\begin{equation}
\Delta_{IR}=\frac{a_2}{a_2-1}d, \ \ \Delta_{UV}=\frac{d}{a_2-a_1}.
\end{equation}
We should impose $\Delta_{IR}>d$ and $d>\Delta_{UV}>\frac{d}{2}$, which implies $a_2>1$ and $1<a_2-a_1<2$. The last condition is not possible for integer numbers, so $c_2$ is always non-zero. Note that if we allow $\Delta_{UV}=\frac{d}{2}$, then $a_2-a_1=2$ can be satisfied and there is a discrete set of models with $c_2=0$.

\subsection{The scalar mode}
The Heun function \eqref{scalarSol} can be reduced to the hypergeometric function when
\begin{equation}
    \alpha=\tau=\gamma=0,
\end{equation}
in the scalar equation \eqref{scalarEq}. That happens when c is tuned to
\begin{equation}\label{cTunedS}
    c=  \frac{
    3( d-2) (d - \Dir)   -(d+6-4 \Dir)\Duv
    }
    {
    (d- 2)(d-\Dir )            -(d+2-2\Dir)\Duv
    }.
\end{equation}
Upon changing variables to
\begin{equation}
  z\equiv\frac{v-1}{b-1},
\end{equation}
the scalar equation \eqref{scalarEq} then reduces to the hypergeometric equation
\begin{equation}
  z(1-z)\partial_z^2Y+(j_2-\omega z)\partial_z Y +2(\omega+1)Y=0.
\end{equation}
There are two independent solutions to this equation
\begin{equation}
  Y= \,_2F_1\left[ -2,\omega+1;j_2;z  \right] , \qquad  z^{1-j_2}\,_2F_1\left[ -1-j_2,2+\omega-j_2;2-j_2;z  \right].
\end{equation}
The first one is identical to $\frac{W(v)}{v^{1-c}}$ which results in the constant solution for the function $G$. The second, non trivial, solution for the function $G$ is therefore
\begin{equation}\label{reducedScalarSol}
  G = \frac{\left[(b-1)z+1\right]^{1-c}}{W(z)} z^{1-j_2}\,_2F_1\left[ -1-j_2,2+\omega-j_2;2-j_2;z  \right].
\end{equation}

The reduced solution can now be expanded around each of the fixed points. Its expansion around the boundary $v=1$ ($z=0$) is
\begin{equation}
  G = z^{1-j_2} \left[-\frac{(c-3) (c-2) (c-1) }{b (c-3)-(c-1)}+\mathcal{O}(z)\right].
\end{equation}
The expansion around the horizon $v=b$ ($z=1$) is
\begin{eqnarray}\label{RedScalarSolHor}
   G &=&
    \left [
    \frac{(c-3) (c-2) (c-1) \pi  \text{Csc}\left[\pi  \left(\omega -j_2\right)\right] \Gamma\left(2-j_2\right) \left(\omega -j_2\right)}{2 b \left[b (c-1)-(c-3)\right] \Gamma(-\omega ) \Gamma\left(1+\omega -j_2\right)}
   + \mathcal{O}(z-1)
   \right]
   \\
   &+&
  \nonumber (z-1)^{j_2 -\omega+1 } \left[
  \frac{(c-3) (c-2) (c-1) e^{i \pi  \left(\omega -j_2\right)} j_2 \left(j_2^2-1\right)}
  {        b \left[b (c-1)-(c-3) \right]  \left(\omega -j_2\right) \left(\omega +1 -j_2\right)  \left(\omega -1 -j_2\right)          }
  +\mathcal{O}(z-1)\right].
\end{eqnarray}
The expansion around the boundary does not contain a constant piece and therefore the mixing term is nothing but the constant piece in \eqref{RedScalarSolHor}
\begin{equation}
  c_2 =     \frac{(c-3) (c-2) (c-1) \pi  \text{Csc}\left[\pi  \left(\omega -j_2\right)\right] \Gamma\left(2-j_2\right) \left(\omega -j_2\right)}{2 b \left[b (c-1)-(c-3)\right] \Gamma(-\omega ) \Gamma\left(1+\omega -j_2\right)}.
\end{equation}
Note that the form of the superpotential \eqref{superpotential} implies that $c\neq 3,2,1$. Then, $c_2$ will vanish only if $\omega$ is a non-negative integer. This leads to the condition
\begin{equation}
\Delta_{IR}=\frac{d(d-2) -\Delta_{UV} (d (\omega +1)+2)}{d-2-(\omega+2) \Delta_{UV}}.
\end{equation}
We should impose that $\Delta_{IR}>d$ and note that, since $\Delta_{UV}>\frac{d}{2}$, the denominator is negative. Therefore, we get the condition $d<2$, which implies that $c_2$ is non-zero.

\subsection{Overlap and validity of the solutions}
In the last two subsections our strategy was to study each of the sectors (scalar and tensor) separately. Using a smart choice of the c-parameter we have reduced the solution to a simpler function. The question is whether there is an overlap between the two different choices of the c-parameter \eqref{cTunedT} and \eqref{cTunedS}, such that the reduced solutions in the two sectors exist at the same time. The answer to this question is yes. When the following relation between $\Duv$ and $\Dir$
\begin{equation}
    \Dir=d\frac{d(d-2)+ \Duv(d-4)}{d(d-2)-2 \Duv},
\end{equation}
holds, the two choices \eqref{cTunedT} and \eqref{cTunedS} for the c-parameter coincides.

As we mentioned before, the specific choice of the c-parameter may not necessarily be compatible with the requirements on the (super)potential.
For the tensor sector we have checked explicitly, in various number of dimensions ($d=2-8$), that there is a range of parameters for $\Duv,\Dir$ which is compatible with the solution \eqref{cTunedT}.

For the scalar case we can make a similar statement for dimensions $d=5-8$. In $d=3$ and $d=4$ spacetime dimensions, the solution \eqref{cTunedS} is compatible only with $\lambda_{UV}<\frac{d}{2}$, which corresponds to explicit breaking of the conformal symmetry.
In $d=2$ spacetime dimensions there is no range of parameters which is compatible with the solution \eqref{cTunedS}.
However, in the next section we study another example in $d=4$ dimensions, corresponding to the spontaneous breaking case, and in appendix \ref{2dExample} we study an example in $d=2$ dimensions.

\section{Example in 4D}

\begin{figure}
  \center
  \includegraphics[scale=0.88]{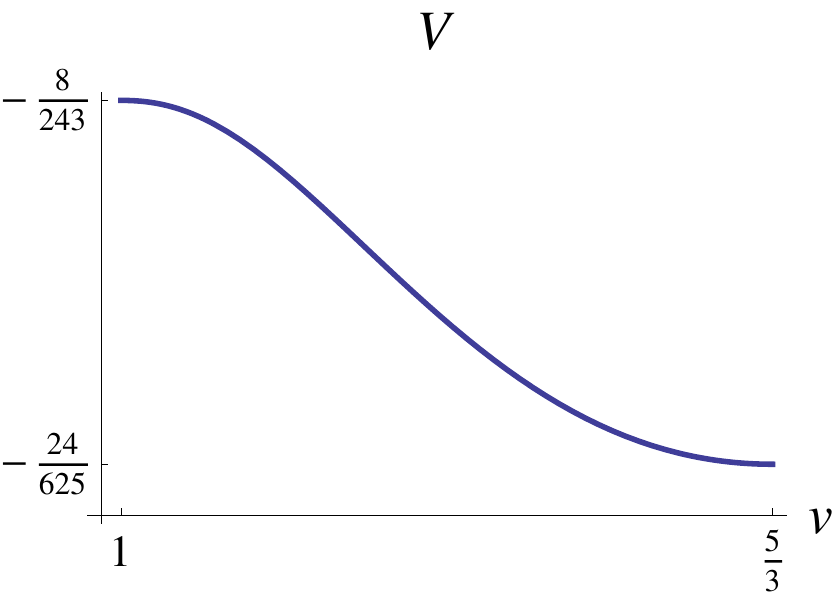}\quad \includegraphics[scale=0.84]{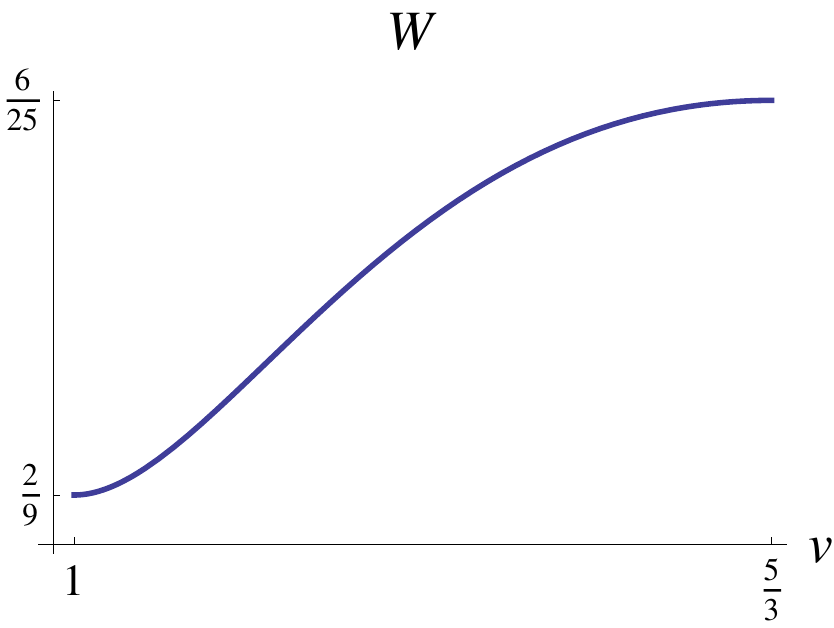}\\
  \caption{The potential and superpotential for example \eqref{35example}, as functions of the variable $v$.}
	\label{vpotential}
\end{figure}

In this section we study a concrete example in four spacetime dimensions where we can solve for both the tensor and the scalar modes.
The superpotential is of the form \eqref{superpotential} with the following set of parameters
\begin{eqnarray}\label{35example}
  \nonumber \Duv &=& 3 ,\\
  \Dir &=& 5 ,\\
  \nonumber c &=& 4.
\end{eqnarray}
The potential and superpotential in this case are depicted in figure \ref{vpotential}.

\subsection{The tensor mode}
For this concrete choice of the parameters, the solution \eqref{tensorSol} to equation \eqref{eqtensor} reduces to
\begin{equation}
   h = \frac{72 (v-1)^{\frac{7}{3}} v^{\frac{5}{3}}}{(3v-5)^5}   \left[
   v \frac{3 v+5}{3v-5}  \, _{2}F_{1}\left(2,6;\frac{17}{3};\frac{2 v}{5-3 v}\right)
   -\frac{7}{9}  \frac{6 v-11}{v-1}  \, _{2}F_{1}\left(1,5;\frac{14}{3};\frac{2 v}{5-3 v}\right)
   \right],
\end{equation}
where $_{2}F_{1}$ is the hypergeometric function.

The expansion around $v=1$
\begin{eqnarray}\label{boundaryExpansion}
  h &=& -\frac{1078 (-1)^{\frac{2}{3}} \left(\frac{2}{5}\right)^{\frac{1}{3}} \pi }{675 \sqrt{3}} + \frac{385}{16} (v-1)^{\frac{4}{3}}  \left( 1+\frac{14 (v-1)}{3} +O(v-1)^2  \right).
\end{eqnarray}
The expansion around $v=b$
\begin{eqnarray}
  \nonumber h &=& \frac{770 2^{\frac{1}{3}} 5^{\frac{2}{3}}}{2187 \left(v-\frac{5}{3}\right)^4}+\frac{308 2^{\frac{1}{3}} 5^{\frac{2}{3}}}{729 \left(v-\frac{5}{3}\right)^3}-\frac{77 \left(\frac{2}{5}\right)^{\frac{1}{3}}}{243 \left(v-\frac{5}{3}\right)^2}+\frac{847 \left(\frac{2}{5}\right)^{\frac{1}{3}}}{1215 \left(v-\frac{5}{3}\right)}+ \\
  \nonumber&+& \frac{77 \left(197+28 \left(\sqrt{3}-12 i\right) \pi -252 \ln(3)-168 \ln(10)+168 \ln(9 v-15)\right)}{8100 2^{\frac{2}{3}} 5^{\frac{1}{3}}} \\
  &+& \frac{6391 \left(v-\frac{5}{3}\right)}{3375 \left(2^{\frac{2}{3}} 5^{\frac{1}{3}}\right)}  \left[ -1 + \frac{1013}{1660}\left(v-\frac{5}{3}\right) +O\left(v-\frac{5}{3}\right)^2 \right].
\end{eqnarray}

In order to get the usual expansion of the solution around the boundary we need to subtract from the solution the constant piece appearing in \eqref{boundaryExpansion}. Then, in the expansion around the $IR$ fixed point we get a constant contribution which will induce a mixing with the constant solution
\begin{eqnarray}
  c_2 &=& -\frac{77 \left(-197+28 \left(6 i+\sqrt{3}\right) \pi +252 \ln3+168 \ln10\right)}{8100 2^{\frac{2}{3}} 5^{\frac{1}{3}}},
\end{eqnarray}
different than zero. This constant piece will mix the two solutions for $h$ at zero momentum.

\subsection{The scalar mode}
For the same example, the scalar equation reduces to
\begin{eqnarray}\label{eqscalar}
  \partial_v ^2 G + \left(  -\frac{2 \left(100-300 v+333 v^2-216 v^3+81 v^4\right)}{3 v \left(5-8 v+3 v^2\right) \left(5-12 v+9 v^2\right)} \right) \partial_v  G   &=& 0,
\end{eqnarray}
for which the solution is
\begin{equation}
    G = \frac{(v-1)^{1/3} v^{2/3} \left[400+9 v (9 (v-4) (v-1) v-80)-80 (5+3 v (3 v-4)) _{2}F_{1}\left(1,1,\frac{5}{3},v\right)\right]}{81 (5+3 v (3v-4))}.
\end{equation}

The expansion of the solution around $v=1$ is
\begin{equation}\label{boundaryExpansionScalar}
    G = \frac{320 (-1)^{\frac{2}{3}} \pi }{243 \sqrt{3}}  + \frac{3}{2} (v-1)^{\frac{4}{3}} \left[ -1+\frac{94 (v-1)}{21} +O(v-1)^2  \right].
\end{equation}
The expansion of the solution around $v=b$ is
\begin{equation}
   G= -(1.28992-2.06854 i) +0.230252 \left(v-\frac{5}{3}\right)^4  \left[ 1 -1.2 \left(v-\frac{5}{3}\right)+O \left( v-\frac{5}{3} \right) ^2 \right].
\end{equation}
Subtracting the constant piece from the expansion around the boundary \eqref{boundaryExpansionScalar} we get in the $IR$ expansion a non-zero constant contribution
\begin{equation}
    c_2 = -0.0956492.
\end{equation}

\section{Explicit breaking}

Our discussion so far concentrated on the spontaneous breaking case, where the leading singularity in the two-point function is that of the massless dilaton pole. When conformal symmetry is explicitly broken we do not expect a massless pole in the propagator. In this section we explain how our results agree with this expectation.

The expansion of the scalar bulk fluctuation near the boundary takes the form
\begin{equation}\label{Bexpansion}
    \varphi = \varphi_b \phi_0^{\frac{d}{\lambda_{UV}}-1}(1+\dots)  +\varphi_n \phi_0(1+\dots).
\end{equation}
The spontaneous breaking case corresponds to $\lambda_{UV}=\Duv>\frac{d}{2}$. In that case the second solution in \eqref{Bexpansion} goes faster to zero near the boundary ($\phi_0=0$) and therefore $\varphi_n$ corresponds to the VEV of the dual operator and $\varphi_b$ to its source. The two-point function is proportional to the ratio
\begin{equation}
     \left<\mathcal{O}\mathcal{O}\right>_{spontaneous}\sim\frac{\varphi_n}{\varphi_b}.
\end{equation}
The explicit breaking case corresponds to $\lambda_{UV}=d-\Duv<\frac{d}{2}$. In that case the first solution in \eqref{Bexpansion} goes faster to zero near the boundary and the two solutions change their roles. $\varphi_n$ then becomes the source for the dual operator. The two-point function in this case is therefore given by the inverse of the previous case
\begin{equation}
     \left<\mathcal{O}\mathcal{O}\right>_{explicit}\sim\frac{\varphi_b}{\varphi_n}.
\end{equation}
Therefore, the correlator in the explicit breaking case will be the inverse of \eqref{CorrSponS}
\begin{eqnarray}
    \left< \mathcal{O}\mathcal{O} \right>_{explicit} &\sim&  Q^2+\dots+Q^{2\Dir-d}+\dots \quad.
\end{eqnarray}
The dots contain analytic powers of the momentum and subleading non-analytic terms. We interpret the analytic terms as contact terms in the field theory. That result agrees with the expectation in field theory: the scalar correlator is proportional to $Q^{2\Dir-d}$ which is the right behavior for the two-point function of an operator of dimension $\Dir$ in the conformal field theory at the $IR$ fixed point.

The results for the correlators involving tensor perturbations, $\left<T^{\mu\nu}\mathcal{O}\right>$ and $\left<T^{\mu\nu}T^{\alpha\beta}\right>$, in the explicit breaking case will require a more careful analysis since local counterterms will change \eqref{mixedCorrelator} and \eqref{correlators1}. We do not expect to have a massless pole in these correlators in the explicit breaking case.

\section{Concluding remarks and future directions}

In this paper we have studied holographic duals to theories with spontaneous breaking of conformal symmetry. A subtlety in the low-momentum expansion, which was pointed out by Bajc and Lugo \cite{Bajc:2013wha}, modifies some of our previous results \cite{Hoyos:2012xc}. As described in section \ref{mixing}, the origin of this subtlety lies in the fact that the normalizable and the non-normalizable modes mix. Each of these modes near the boundary is a superposition of the modes in the near horizon region.

Our main result in this paper is the scalar-scalar and tensor-tensor two-point functions \eqref{CorrSponT}-\eqref{CorrSponS}
\begin{eqnarray}
   \label{corrConc1} \left< T^{\mu\nu} T^{\alpha\beta} \right> &\sim& \Pi^{\mu\nu,\alpha\beta} \frac{1}{Q^2+\dots+Q^{d}\ln Q^2+\dots} ,\\
   \label{corrConc2} \left< \mathcal{O}\mathcal{O} \right> &\sim&  \frac{1}{Q^2+\dots+Q^{2\Delta_{IR}-d}+\dots},
\end{eqnarray}
which are affected by the mixing.
The dots after the $Q^2$ terms stand for higher order, integer powers, corrections in momentum, while the dots after the other terms stand for higher order, non-integer powers, corrections in momentum.
The correlator $\left< T^{\mu\nu}\mathcal{O} \right>$ is not affected by the mixing and hence our previous result \cite{Hoyos:2012xc} still holds.

\begin{figure}
  \center
  \includegraphics[scale=0.4]{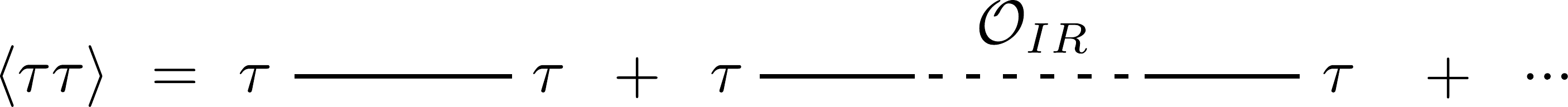} \\[0.5cm] {figure 3.1} \\[1cm]
  \includegraphics[scale=0.4]{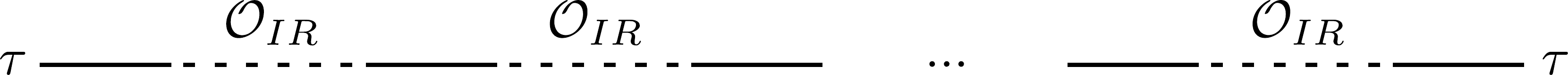} \\[0.5cm] {figure 3.2} \\[0.5cm]
  \caption{The dilaton interacting with the $CFT_{IR}$. The theory is described by the Lagrangian \eqref{lagrangian}. The upper figure describes the infinite set of diagrams which contribute to the dilaton two-point function. The lower figure describes a typical diagram in the sum, with $n$ intermediate $\mathcal{O}_{IR}$ states.}
	\label{dilatonInt}
\end{figure}

Let us interpret the results \eqref{corrConc1}-\eqref{corrConc2}. The theory in the $IR$ contains a dilaton which is coupled to a conformal field theory. The Lagrangian is therefore of the form
\begin{equation}\label{lagrangian}
  \mathcal{L} =   \frac{1}{2} (\partial \tau)^2 +  \mathcal{L}_{IR}  + g\tau \mathcal{O}_{IR}    + \dots \quad,
\end{equation}
where "$\dots$" represent higher order interactions.
$\tau$ is the canonically normalized dilaton, $\mathcal{O}_{IR}$ is an operator of dimension $\Dir$ in the $IR$ CFT and $g$ is the coupling between them.
It is important to note that $\mathcal{O}_{IR}$ is different than the operator $\mathcal{O}$ which is of dimension $\Duv$.
In order to calculate the two-point function of the dilaton one needs to sum over infinite number of diagrams, containing intermediate $\mathcal{O}_{IR}$ states, as described in figure \ref{dilatonInt}.
A diagram that contains $n$ intermediate $\mathcal{O}_{IR}$ states is equal to
\begin{equation}
  \mathcal{A}_n = \frac{1}{Q^2}  \left[    g^2 \frac{1}{Q^2} \left< \mathcal{O}_{IR}\mathcal{O}_{IR} \right>   \right]^n .
\end{equation}
As usual, the diagrams can be summed as a geoemtric series, leading to the result
\begin{equation}
  \left< \tau \tau  \right> =  \frac{1} { Q^2-   g^2  \left< \mathcal{O}_{IR}\mathcal{O}_{IR} \right>   }  \simeq
  \frac{1} { Q^2-   g^2  Q^{2\Dir-d}    } .
\end{equation}
The dilaton overlaps with the operator $\mathcal{O}$ \eqref{relation} and therefore $\left< \tau \tau  \right> \sim \left< \mathcal{O}\mathcal{O} \right>$.
From the point of view of the $IR$ theory \eqref{lagrangian} the higher order corrections in momentum in \eqref{corrConc2} are due to higher order interactions.
This analysis explains the structure of the correlator \eqref{corrConc2}. A similar argument explains also \eqref{corrConc1}. As the dilaton also interacts with the energy-momentum tensor, the result for the tensor-tensor correlator will be similar, with $\left< \mathcal{O}_{IR}\mathcal{O}_{IR} \right>$ replaced by $\left< T^{TT}_{IR}T^{TT}_{IR} \right>$. The superscript $^{TT}$ stands for "transverse traceless" and
\begin{equation}
  \left< T^{TT}_{IR}T^{TT}_{IR} \right> \simeq  Q^d \ln Q^2.
\end{equation}
The $\ln Q^2$ term is present only in even dimensional theories.

An interesting feature of the leading pole singularity in \eqref{corrConc2} is that it does not depend strongly on the boundary conditions. While the non-analytic term $Q^{2\Dir-d}$ in \eqref{corrConc2} is a direct consequence of the boundary conditions on the horizon, the massless pole is more kinematical in nature, by which we mean that it is robust to changes in the boundary conditions. The definition of the holographic dilaton seems to be closely related to the mixing between the normalizable and non-normalizable modes, which is a kinematical, rather than dynamical, feature. In particular, the dilaton mode can be present even in singular geometries, as has been observed in some examples of holographic duals to ${\cal N}=4$ SYM in the Coulomb branch \cite{Freedman:1999gk,Mueck:2001cy}. We should remark that contrary to the ${\cal N}=4$ SYM example, our models do not in general belong to a consistent truncation of supergravity  (SUGRA) and as a consequence the holographic dual is unkown. In general we expect that in a realistic theory the operator spectrum contains more than a single scalar operator, and correspondingly more than one scalar field in the holographic dual. The class of models where our analysis apply are those where the relevant operator that gets an expectation value does not mix with other operators along the RG flow. This is possible when the dual field in the truncated gravitational model is decoupled from other scalar fields. Such examples exist in SUGRA models but there is no known case where the geometry close to the horizon is $AdS$, instead some kind of singularity is present. We comment more extensively on this feature in our previous paper \cite{Hoyos:2012xc}.

There are many future directions that will be interesting to pursue:
\begin{enumerate}
  \item It would be interesting to go beyond the low momentum limit and calculate the correlators for any momentum. In general that is not possible but results can be obtained for specific superpotentials, or by using other approximations, such as WKB.
  \item We have considered asymptotically $AdS$ spaces with an $AdS$ horizon in the bulk. However, one can repeat the same steps for general asymptotically $AdS$ spaces with different kinds of horizons in the bulk. The boundary conditions and the infrared structure of the theory will depend on the form of the horizon. It may be possible to study the general structure of this theories along the lines of \cite{Nickel:2010pr} and using the effective field theory approach \cite{Cheung:2007st}.
  \item It would be interesting to relate the exact solution in two dimensions we found in section \ref{exact2D} to the Zamolodchikov's c-theorem \cite{Zamolodchikov:1986gt}.
  \item A similar analysis to the one presented in this paper can be done for fluctuations in \emph{de Sitter} space \cite{Kol:2013msa}. The correlators of a theory dual to a flow between two dS space of different Hubble constant can be calculated. It would also be interesting to see the effect of the mixing between the normalizable and the non-normalizable on the inflationary correlators \cite{Maldacena:2002vr}. Possible effects could be the enhancement of superhorizon perturbations \cite{Leach:2001zf} or suppression of non-gaussianities.
  \item The two-dimensional case showed agreement with anomaly matching arguments, it would be interesting to study four-point functions of the dilaton and try to identify the term associated to the $a$-anomaly in four dimensions.
\end{enumerate}

\acknowledgments
We thank Borut Bajc and Adrian R. Lugo for pointing out to us the subtlety in the matching procedure and for interesting correspondence.
U.K. is supported by the Lev Zion fellowship of the Council for Higher Education (Israel).
This work was supported in part by the Israel Science Foundation (grant number 1468/06).

\appendix

\section{More on the solutions}\label{ExactSol}

In this appendix we provide the details of the solutions for the model presented in section \ref{model}. We express the parameters of the equations of motion with the following set of variables describing the model $(\beta,b,c)$. The translation to the physical set of parameters $\Duv,\Dir,c$ is done using the translation \eqref{btrans}-\eqref{betatrans}.

\subsection{The tensor mode}
The equation of motion for the tensor mode is given by \eqref{eqtensor} with the following parameters
\begin{eqnarray}
  \omega &=& \frac{d}{(c-3) (d-1) \beta ^2}  -(c-4) ,\\
  \gamma+\tau &=& \frac{(b+1) (c-3) }{  (c-2) }
  \frac{(d-1)(c-3) (c-2)  \beta ^2  -d }{(d-1)(c-4) (c-3) \beta ^2  -d} ,\\
  \gamma-\tau &=& \frac{
  \sqrt{(c-3) (c-1)x_T}
  }
  {
  (c-2) (c-1) \left[(c-4) (c-3) (d-1) \beta ^2-d \right].
  }
\end{eqnarray}
where
\begin{eqnarray}
 \nonumber x_T &=&
  (b+1)^2 (c-3) (c-1) \left[(c-3) (c-2) (d-1) \beta ^2 -d \right]^2  \\
  &&
  -4 b (c-2)^2 \left[(c-4) (c-3) (d-1) \beta ^2-d\right] \left[(c-2) (c-1) (d-1) \beta ^2-d\right].
\end{eqnarray}

\subsection{The scalar mode}
The equation of motion for the scalar mode is given by \eqref{scalarEq} with the following parameters

\begin{eqnarray}
  \omega &=& -\frac{d-2}{  (d-1) (c-3)  \beta ^2}  - c ,\\
  \alpha &=&   \frac{(b+1) (c-3) }{2 (c-2) }
  \frac{(d-2)+(d-1)(c-2) (c-1) \beta ^2}{(d-2)+(d-1) (c-3)(c-1)\beta ^2} ,\\
  \gamma+\tau &=& \frac{(b+1) (c-3) }{(c-2) }
  \frac{(d-2)+(d-1) (c-1)(c-2) \beta ^2}{(d-2)+(d-1)(c-3) c \beta ^2},\\
  \gamma - \tau &=&  \frac{
\sqrt{(c-3) (c-1) \left[(d-2)+(d-1)(c-2) (c-1) \beta ^2\right] x_S}
}
{
(c-2) (c-1) \left[(d-2)+(d-1)(c-3) c \beta ^2\right]
},
\end{eqnarray}
where
\begin{eqnarray}
 \nonumber x_S &=& (d-2) \left[ b (c-3)-(c-1) \right] \left[ b (c-1) -(c-3) \right] \\
   &+& (d-1)(c-3) (c-2)  \beta ^2 \left[  (b+1)^2    +   c(c-2) (b-1)^2  \right].
\end{eqnarray}

\section{Example in 2D}\label{2dExample}

In this appendix we study an example in two spacetime dimensions.
The scalar mode can be solved exactly for a general superpotential (see section \ref{exact2D}). In order to demonstrate how the mixing works in the tensor sector we study an explicit example.
The superpotential is of the form
\begin{equation}\label{superpotential2D}
  W = \frac{1}{3}v^3  -\frac{b+1}{2} v^2      +b v ,
\end{equation}
(where the dimensionful parameter was set to one, as in section \ref{model}).
The derivative of the superpotential is
\begin{equation}
  \partial_v W = (v-1)(v-b).
\end{equation}
The superpotential therefore have two critical points, one at $v=1$ and the other at $v=b$.

The two parameters of the model are $\beta$ and $b$. The dimensions of the $UV$ and $IR$ operators are
\begin{eqnarray}
  \Duv &=& -\frac{6 (b-1) (d-1) \beta ^2}{3 b-1} ,\\
  \Dir &=& d+\frac{6 (b-1) (d-1) \beta ^2}{b-3}.
\end{eqnarray}
We can then use the dimensions $\Duv,\Dir$ as the two parameters of the model using the invert relations
\begin{eqnarray}
  b &=& \frac{3 (\Dir-d)+\Duv}{\Dir-d+3 \Duv} ,\\
  \beta^2 &=& \frac{2 (\Dir-d) \Duv}{3 (d-1) (d-\Dir+\Duv)}.
\end{eqnarray}

\subsection{The tensor mode}

The equation of motion for the tensor perturbation is of the form \eqref{eqtensor} with the following parameters
\begin{eqnarray}
  \omega &=& 4-\frac{d}{3 (d-1) \beta ^2} ,\\
  \tau+\gamma &=& \frac{3 (b+1)}{2}   \frac{ 6 (d-1) \beta ^2-d }{12 (d-1) \beta ^2- d} ,\\
  \tau-\gamma &=&   \frac{  \sqrt{3 y_T  }}  {24 (d-1) \beta ^2- 2d},
\end{eqnarray}
where
\begin{equation}
    y_T=(b-3) (3 b-1) d^2
  -4 \left[9+b (9 b-38)\right] (d-1) d \beta ^2
  +12 \left[9+b (9 b-14)\right] (d-1)^2 \beta ^4.
\end{equation}

The solution is again given by the Appell hypergeometric function \eqref{tensorSol} with
\begin{eqnarray}
  a_1 &=& \frac{3 d}{2\Duv}   \frac{d-\Dir+\Duv}{\Dir-d}     - 1  ,\\
  a_2 &=& \frac{\Dir}{\Dir-d} ,\\
  a_3 &=& 1-\frac{d}{\Duv}.
\end{eqnarray}
In order to reduce \eqref{tensorSol} to the incomplete Euler beta function we want to tune one of the parameters of the model such that
\begin{equation}
  a_1+1=a_2+a_3.
\end{equation}
That happens when
\begin{equation}\label{tuneM}
  \Dir = d\frac{d+5 \Duv}{d+4 \Duv}.
\end{equation}
The solution is then given again by equations \eqref{reducedTensorSol}-\eqref{mixCoeffTensor}. However, when plugging \eqref{tuneM} in the mixing coefficient \eqref{mixCoeffTensor} we find that it vanishes
\begin{equation}
  c_2=0.
\end{equation}
We see that in some cases and for some specific choice of the parameters the mixing constant can vanish.

\end{document}